\documentclass[11pt,twoside]{article}

\usepackage{asp2004}
\usepackage{epsf}
\usepackage{psfig}
\usepackage{lscape}
\usepackage{graphicx}

\begin{document}
\title{Z-machine science other than CO -- scientific and technical prospects
for very wide band-width radio and (sub)millimeter-wavelength spectroscopy}   
\author{Karl M. Menten}   
\affil{Max-Planck-Institut f\"ur Radioastronomie}    

\begin{abstract} 
While clearly the main scientific targets of ''z-machines"
will be redshifted lines of carbon monoxide, there also
exist other interesting applications.
Here scientific and technological aspects of observing lines from CO and other species
at high redshift and in the local universe are discussed as are
the limitations of such efforts and prospects for
the future.
\end{abstract}

\section{Introduction}
At present, several concepts for ''z-machines" -- detector/spectrometer
combinations covering wide spectral ranges -- are discussed for different
wavelength regimes. Systems presented at this workshop are
summarized in Table 1.
\begin{table}[!ht]
\caption{Z-machines currently under discussion}
\smallskip
\begin{center}
{\small
\begin{tabular}{ccccc}
\tableline
\noalign{\smallskip}
System & Frequency range & Velocity Resolution$^{\rm a}$ & Telescope$^{\rm b}$ \\
\noalign{\smallskip}
\tableline
\noalign{\smallskip}
Zpectrometer$^{\rm c}$           & 28.5 -- 34.5 GHz   & 150 km~s$^{-1}$ & GBT$^{\rm d}$\\
RSR$^{\rm e}$			& 74 -- 110 GHz      & 100 km~s$^{-1}$ & LMT$^{\rm f}$\\
Zspec$^{\rm g}$                  & 185 -- 293 GHz     & 800 km~s$^{-1}$ & (sub)mm telescopes\\
ZEUS$^{\rm h}$			& various submm bands& 300 km~s$^{-1}$ & submm telescopes\\
\noalign{\smallskip}
\tableline
\end{tabular}
}
\end{center}
$^{\rm a}$Velocity resolutions vary over/with band.
$^{\rm b}$All instruments can be adapted to
various telescopes suitable for their wavelength range.
$^{\rm c}$Harris et al.  (these proceedings).
$^{\rm d}$Green Bank Telescope
$^{\rm e}$Redshift Search Receiver; see Erickson et al. (these proceedings).
$^{\rm f}$Large Millimeter Telescope -- Gran Telescopio Milim\'etrico
$^{\rm g}$Glenn et al. (these proceedings).
$^{\rm h}$The Redshift(z) and Early Universe Spectrometer; see Stacey et al. (these proceedings).
\end{table}
Any search for extremely weak radio or (sub)millimeter (submm)
high-$z$ line emission requires considerable investment
in observing time. We therefore start (in \S \ref{highz})
by summarizing high-$z$ cm and mm spectroscopy. In particular we discuss the pros
and cons of building z-machines in the centimeter range, i.e. to search
for low rotational quantum number ($J_{\rm upper}=1$ or 2) carbon monoxide (CO) and
(in \S \ref{hcn}) hydrogen cyanide (HCN) transitions.

Largely because of their relatively coarse velocity resolution,
for  all the z-machines under
discussion meaningful observations of the local universe
are restricted to extragalactic astronomy, i.e. spectroscopic
observations of nearby galaxies.
In \S \ref{ngc253} we see
that there is indeed very interesting science to be explored.

Limitations to attempts at detecting very weak, broad line emission
are discussed in \S \ref{baselines} together with methods to overcome them.
We point out the need for
wide instantaneous bandwidth, high spectral resolution (= large channel number)
detector/spectrometer combinations.

In \S \ref{ffts} we present a novel spectrometer concept, the Fast Fourier
Transform Spectrometer and we argue that, rather than building low resolution
spectrometers, a very wide-band receiver together with such a very high resolution
spectrometer would allow, in addition to extragalactic science, fantastic galactic
astronomy,
examples of which are briefly presented in \S \ref{galactic}

\section{\label{highz}High-$z$ z-machine science}
\subsection{CO}
Over the last few years, redshifts of many high-$z$ submillimeter dust
emission (``SCUBA'') sources have been measured
by optical spectroscopy using large telescopes (Chapman et al. 2003, 2005)
allowing searches for mm-wavelength lines (mostly from CO), which frequently are
successful (see contributions by Cox and Wei\ss\ to these proceedings).

In some cases the redshifts delivered by
optical spectroscopy can be different by the equivalents of hundreds of km~s$^{-1}$
from the CO redshifts. This presents a
problem for the relatively narrow band-passes of the most prolific telescope
for detecting such sources, the IRAM Plateau de Bure Interferometer (PdBI). However, this
situation will be remedied in the near future, by a refurbishing with
broader intermediate frequency (IF) band receivers and a broader band correlator
and wouldn't motivate the need for a z-machine.

Despite much progress, about a third of all SCUBA sources remain unidentified
at other wavelengths. Since their non-discovery at optical or even near-infrared
wavelengths may mean that these objects have immense amounts of dust and, thus,
extremely high star formation rates, determining their redshifts
and, thus, their luminosities,
is of great interest.  For this (sub)mm observations are invaluable as
optical/ultraviolet observations are heavily biased by dust obscuration.

For such objects  photometric methods using the radio-to-infrared
relation result in crude redshift estimates (Carilli \&\ Yun 1999, 2000;
Hughes et al. 2002; Aretxaga et al. 2003 and these proceedings).
Obtaining accurate redshifts and, thus, CO luminosities
are main motivations for z-machines.

Z-machines in the centimeter range, in particular in the  $K$- and $Ka$-bands (18 -- 26.5 and
26.5 -- 38 GHz, respectively)
are  partially motivated by the desire to probe
CO distributions similar to those that constitute the large-scale molecular
gas in ``normal'' galaxies  ($J_{\rm upper} \sim 3$) . Over most of the Milky Way
the $J = 3-2$ transition has the highest emissivity, except for the central few degrees
in which the $J = 4-3$ and $5-4$ lines dominate (Bennett et al. 1994).
This extended gas is significantly cooler ($\sim10$ -- 20 K) than
the $> 60$ K warm interstellar medium (ISM) probed
by the mid- and higher $J$ lines ($J = 4, 5, 6, 7 \ldots$)
that are shifted to mm-wavelengths and sample starburst regions and/or the vicinities
of active galactic nuclei  (AGN).

Recently, the CO $J=1 - 0$ line was discovered with the Green Bank Telescope (GBT)
and the
Effelsberg 100 m telescope toward three $z > 4$ objects,
the QSOs BR 1202-0725 (z=4.7), PSS J2322+1944 (z=4.1), and APM 08279+5255 (z=3.9)
(Riechers et al.
2006 and these proceedings). These observations
required dozens of hours of telescope time.
Comparing their $J=1 - 0$ results with those obtained for higher-$J$ lines,
Riechers et al. conclude that all are consistent with a single (warm) gas component
that well explains the observations for a wide range of $J$  (up to $J_{\rm upper} = 11$;
see Wei\ss\ et al. these proceedings and in preparation).
These result seem to discourage $J = 1 - 0 $
searches, which are quite expensive compared to searches for higher-$J$
transitions\footnote{The line flux density increases as $\nu^2$ (i.e., $J^2$), whereas the
detection sensitivity for the higher-$J$ mm observations is only a few times worse than for
$J=1-0$ cm observations}. However, we
note that this conclusion so far only holds for the few sources intensely studied:
the three sources mentioned above and the radio galaxy
4C 60.07 (Greve et al. 2004),
which all contain strong AGN and for the Extremely Red Object (ERO) J164502+4626.4,
which possibly also contains an AGN
(Greve et al. 2003).
The situation might be different for non-AGN dominated
submillimeter galaxies.

We also note that, Australia Telescope Compact Array (ATCA) results for the
radio galaxy TN J0924-2201 seem to indicate that the CO $J=1-0$ and $5-4$ emissions
associated with this object
arise from different locations that are separated by $\sim50$ kpc (Klamer et al.
2005).

By far the most powerful instrument to detect and image lower-$J$ CO will be the
Expanded Very Large Array (EVLA)\footnote{http://www.aoc.nrao.edu/evla/}
which will complement higher-$J$ CO observations made with the Atacama Large Millimeter Array
(ALMA). The EVLA will have an 8 GHz IF bandwidth at the  $K$- and $Q-$bands (19 -- 50 GHz).
A first glimpse at the EVLA's promise were recent VLA observations that resulted in the
first resolved CO images a high redshift object, the $z=6.42$  SDSS quasar
J1148+5251 (Walter et al. 2004).
While the VLA today has the required sensitivity, these observations suffer from the woefully
inadequate bandwidth/channel number combination of the ancient VLA correlator.

While the GBT and the EVLA will have comparable effective collecting areas, EVLA
observations will be largely preferable to single dish observations, which will
inevitably be limited by baseline quality (see \S \ref{baselines}).

Given the timelines for the completion of  ALMA and  EVLA (early 2010s), single-dish z-machines
will be interesting for many years in the interim and
it is well worth asking what kind of useful applications they may have other
than high-$z$ CO searches (see \S \ref{ngc253}).

\subsection{\label{hcn}High-redshift emission from molecules other than CO}
Solomon et al. (1992) detected the  HCN $J=1-0$ transition (rest frequency,
$\nu_{\rm rest}$,
near 88.6 GHz) toward five local ultraluminous infrared galaxies (ULIRGS)
and other less luminous systems.
They found an average HCN to CO
luminosity ratio of 1/6, which is much higher than 1/80, the value found in normal spiral
galaxies. Since the critical density, $n_{\rm crit}$, of this HCN line is $\sim10^5$ cm$^{-3}$, much
higher than that of low-$J$ CO lines (hundreds to thousands  cm$^{-3}$),
one conclusion was that the relative percentage of
dense star-forming gas was significantly higher in ultraluminous systems, which have
\textit{bona fide} much higher star formation rates than  normal galaxies.
For their sample Solomon et al. derived correlations between the far infrared (FIR) luminosity and the HCN luminosity
and between the FIR/CO luminosity ratio and the HCN/CO luminosity ratio, which suggest that the
latter is a good indicator for star formation activity. Extrapolating to the even higher
FIR luminosities of high-$z$ sources one would expect the  HCN/CO luminosity ratio to approach
one. This trend seems, however, not to continue to higher luminosities (see below).

It should be noted that these results are for the $J=1-0$ lines of both species.
For linear molecules, $n_{\rm crit} \propto J^3\mu^2$, where the electric dipole
moment, $\mu$, is 0.11 Debyes for CO and 3.0 D for HCN. For line opacities, $\tau > 1$,
$n_{\rm crit}$ is lowered by a factor of
$1\over\tau$. CO excitation modeling of multi-line high-$z$
datasets indicate substantial opacities
for higher-$J$ transitions (many tens to $\sim100$), peaking at $J \sim 6$
(Wei\ss\ et al. 2006 and these proceedings).
For $\tau < 1$ the critical densities of HCN lines would be $\approx 300$
times higher than those of CO lines of identical quantum number.

Given the bandwidths of the planned z-machines, for many values of $z$ there
will be one or more HCN lines in their bands along with CO lines.
So far,  high-$z$  HCN $J=1-0$ emission has been detected only from two sources,
both gravitationally lensed: the
Cloverleaf quasar (H1413+177) at $z=2.56$ (Solomon et al. 2003).
and  IRAS F10214+4724 at $z=2.29$ (Vanden Bout et al. 2004).
Extrapolating to other sources is not straightforward and,
given that, both, the Cloverleaf and IRAS F10214+4724 are amongst
the strongest higher-$z$ CO sources, that the HCN/CO ($J=1-0$)
luminosity ratio (extrapolated for CO from the $3-2$ line)
is very low (0.07 and 0.18, respectively)
and that the HCN lines
are exceedingly weak ($\sim 200$ and $450~\mu$Jy)
the prospects for detecting high-$z$ HCN
are not promising for low $J$ values. Indeed do Carilli et al. (2005)  not obtain
any other clear detection of the $J=1-0$ HCN line in a sensitive search  toward
four other high-redshift far-IR-luminous galaxies.

For higher $J$, on the other hand, at which
CO is easier to detect than at low $J$, the critical densities for HCN
excitation are almost certainly
prohibitive: $n_{\rm crit}$(HCN)  $= 8~10^6$ cm$^{-3}$ and $1~10^8$ cm$^{-3}$
for the $3-2$ and $7-6$ transitions, although these values will
be lowered by the fact that HCN lines will also be optically thick.
Nevertheless,  $n_{\rm crit}$(CO) is only
$5~10^4$ cm$^{-3}$ and $1~10^6$ cm$^{-3}$ for these lines.

To get meaningful input for searches of high-$z$ lines from HCN and other species it
would be important to first observe these lines in local ULIRGSs.
Given the meager observational situation this is particularly important
as signficant variations in relative intensity of the $J=1-0$ lines of CN and HNC
relative to HCN $1-0$ have been reported in a sample of infrared-luminous galaxies
(Aalto et al. 2002) with HCN intensities generally stronger, but frequently
not much stronger, than HNC and CN intensities.

As to molecules other than CO, HCN, HNC, and CN,
as discussed at the
end of \S \ref{ngc253}, available  observations of local starbursts do not give reason for
optimism as to their detectability at high redshift.

\subsection{Carbon (CI and CII)}
Neutral atomic carbon (CI) has two submillimeter-wavelength fine structure
lines:  $^3P_1 - ^3P_0$ ($\nu_{\rm rest}$ near 492 GHz) and
$^3P_2 - ^3P_1$ ($\nu_{\rm rest}$ near 809 GHz). The lower frequency line has now been detected
in three $z = 2.3$ -- 2.5 sources, including the Cloverleaf (Barvainis et al. 1987;
Wei\ss\ et al. 2003). The 809 GHz line so far has only been detected in the latter (Wei\ss\ et al. 2003)
in which both lines have  more than 2 times lower flux density than the
CO $J = 3-2$ line. If, as is likely, in other sources
the flux densities of the CI lines also are lower than those of CO lines, the
former are not  interesting as redshift beacons for z-machines. However, given the frequencies
of the CO $J=4-3$ and $8-7$ lines (461 and 897 GHz, respectively)
any z-machine detecting one of these CO lines
is also likely to detect a CI line for free.

In contrast to the CI lines, which essentially sample the same gas as CO, i.e.,
molecular clouds,
the $158~\mu$m (1900 GHz) $^2P_{3/2} - ^2P_{1/2}$ fine structure line of
ionized carbon (CII) arises from the more extended ISM. It is the most
luminous cooling line
in the ISM of the Milky Way and other low to moderate luminosity galaxies
with FIR luminosities, $L_{\rm FIR}$, $<10^{12} L_\odot$
(e.g., Stacey et al. 1991; Malhotra et al. 2001). In such systems, this line's
luminosity,  $L_{\rm CII}$,
contributes between 0.1 and 1\%\ of  $L_{\rm FIR}$.
For more luminous galaxies (ULIRGs) ISO observations show that
the relative contribution of  $L_{\rm CII}$ to $L_{\rm FIR}$ is about
an order of magnitude smaller (e.g. Luhman et al. 2003; see Maiolino et al. 2005
for a discussion of possible explanations).
For $z>6.0$ the line gets redshifted to 270 GHz, the upper cut-off of the IRAM 30m
telescope's 1.3 mm receiver. Maiolino et al. (2005) made the first high-$z$
detection of this line toward  J1148+5251 at $z=6.42/256.2$ GHz, which would
be within the range of Zspec.

Today (before the ALMA era) it is only possible to observe the CII
line from the ground for $z > 6.0$. This is because doing so requires
the largest collecting area telescopes (IRAM 30m and PdBI) under exceptional
weather conditions. With its much larger collecting area and
far better site, ALMA will make lower-$z$ CII observations possible in higher frequency submm
windows.

Observations of the CII line clearly are of paramount importance
for a characterization of the emitting sources' ISM. However,
as to the potential of CII searches for redshift determination, unless there
are sources with much larger $L_{\rm CII}$/$L_{\rm FIR}$
(and larger $L_{\rm CII}$/$L_{\rm CO}$) ratios than known today
(which is not impossible) it is far less expensive to search for CO lines
at lower frequencies (e.g. with the RSR) for this purpose.

\section{\label{ngc253}Extragalactic interstellar medium chemistry}
Since the late 1980s  Mauersberger \&\ Henkel and
their collaborators have detected (mostly with the IRAM 30m telescope)
a large number of
molecules toward the nuclei of nearby galaxies
Species include CH$_3$OH, CN, C$_2$H, HCN, HNC, HCO$^+$, HC$_3$N, CS,
N$_2$H$^+$, SiO,
HNCO, CH$_3$CCH, CH$_3$CN, SO$_2$, NS, NO, H$_2$CO, and NH$_3$.
These (or a selection of these)
molecules are found toward the nuclear regions of
NGC 253,  IC 342,  NGC 6946, M82, NGC 4945, NGC 6946, Maffei 2, i.e.
mostly starburst galaxies. For many species rare isotopomers are detected
and for some multi-transition studies afford excitation analyses.
Except for NH$_3$, most of the observations were made in the
3 and 2 mm ranges and sampled low-$J$ lines for diatomic and linear species.

The nearby ($D \sim 3$ Mpc) starburst galaxy NGC 253's chemistry was studied in most detail.
Martin et al. (2006) scanned
most of the 2 mm window (129.1 to 175.2 GHz) and discovered 111 emission features
from 25 molecular species.   They find that ``the chemistry of NGC 253 shows
a striking similarity with the chemistry observed toward the Galactic
center molecular clouds, which are thought to be dominated by low-velocity shocks.
This resemblance strongly suggests that the heating in the nuclear environment
of NGC 253 is dominated by the same mechanism as that in the central region of
the Milky Way." Martin et al. also emphasize similarities between the chemistry of
NGC 253 and IC 342 and NGC 4945 and pronounced differences to that of M82.
\begin{figure}[!h]
\includegraphics[scale=0.49,angle=-90]{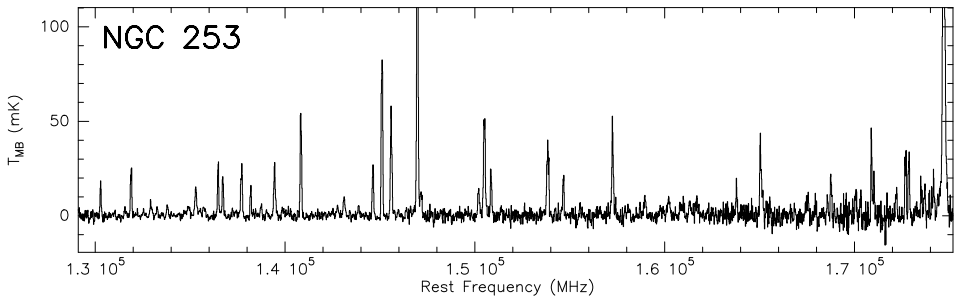}
\caption{The 129.1 to 172.2 GHz range spectrum of the central region of NGC 253.
This spectrum was ''mosaiced'' from a total of 46 individual spectra. The channel spacing
is  69 km~s$^{-1}$.}
\end{figure}
All in all these studies (and studies of our own Galactic center at
much higher physical resolution) provide a fascinating way to probe
lower luminosity versions of the starburst cores of ULIRGs,
local and at high redshift.

Let us check the suitability of the z-machines in Table 1 for studies
such as these.
Martin et al. smoothed their final spectrum to a resolution of 69 km~s$^{-1}$
and even a factor of 2 lower resolution would be adequate to avoid line
blending. Such resolutions could be easily afforded by a 2 or 3 mm
version of the Zpectrometer or by the RSR.
Zspec, on the other hand, would have severe line blending problems due to
its coarse ($\sim 800$ km~s$^{-1}$) resolution.

The observing time requirements of the  NGC 253/IRAM 30m line survey
were completely dictated by the available receiver/spectrometer equipment:
2 SIS receivers with 1 GHz bandwidth each operated in parallel.
HEMT receivers with competitive noise performance will be available soon
at these wavelengths and a single one would cover the whole 46 GHz bandwidth
instantaneously, reducing the
observing time from ca. 50 h to just two.
Similar arguments hold for the 3 mm band already
now.
This means that extremely exciting
science of this kind could be done with the RSR now and in the 2 mm band
with a future HEMT receiver on a large number of and/or weaker sources.

Detecting lines from the molecules observed in NGC 253
in high-$z$ sources seems prohibitive,
however. Just note that the strongest line in the NGC 253 2 mm survey (CS $J=3-2$)
has a peak flux density of ca. 0.5 Jy. It is ca. 40 times weaker than
the CO $J=1-0$ line toward
the same region (spectrum presented by Ott et al. 2005), which
has a flux density of ca. 20 Jy!
\section{\label{baselines}Getting flat baselines}
Given that that many instrumental and atmospheric sources of
error are uncorrelated between its unit telescopes, an
interferometer naturally provides high quality spectral
baselines (provided the passband is calibrated). In contrast,
the baseline quality  of spectra obtained with single dish telescopes
are frequently compromised for a number of reasons such as power
level variations between on and off source scans or reflections by feed structures.
Although the GBT's unblocked aperture should minimize the latter,
baseline problems nevertheless persevere (see Hainline \&\ Blain,
these proceedings) unless special steps are taken.

An ingenious concept consisting of a set of analog lag cross-correlation spectrometers
combined with a multi-channel correlation radiometer
aims to overcome these
problems in the case of the Zpectrometer (Harris et al., these proceedings),
while the RSR employs a dual
beam switching scheme and a Faraday rotation switch allowing very fast
(1 kHz) switching between the beams (Erickson et al.; these proceedings.)

For (sub)millimeter telescopes a breakthrough in baseline quality
(and huge improvements in overall observing quality and efficiency) was
achieved by the installation of ``wobbling'' subreflectors
that ``chop'' between the target and an off position (typically a
few FHWM beam diameters away) on a $> 1$ Hz cycle. Recent extensive
multi-high-$J$ high-$z$ CO line observations with the IRAM 30 m telescope
have resulted in a large  number of spectra whose excellent quality was verified
by comparison with spectra obtained with the PdBI (Wei\ss\ et al. 2005a,b and these proceedings).

Because of their large size and weight, it may not be possible to
wobble secondaries of the large centimeter-wavelength telescopes
and special solutions as those described above may be required. Also,
adaptive, deformable secondaries, an example of which
will be installed at the Effelsberg
telescope soon, should improve overall data quality.

Nevertheless, at mm-wavelengths single-dish chopped observations yield
baselines that are flat over velocity ranges that are larger
than the linewidths of galaxies and very wideband observations
should be possible with HEMT receivers. Special purpose designs
such as the Zpectrometer's, Zspec's and the RSR's certainly
hold great promise. However, these instruments are hardwired to
certain frequency ranges and have a fixed, coarse velocity resolution.

Rather than building special purpose, low resolution
spectrometers, in the following we strongly advocate to do the opposite, i.e. to build
high resolution, universally usable devices that cover very wide bands
and can be employed in \textit{any} wavelength region.
These are just becoming available in the form of Fast Fourier Transform
Spectrometers (see below).

\section{\label{ffts}Fast Fourier Transform Spectrometers}   
Progress in digital electronics has resulted  in
an exciting new spectrometer concept: The Fast Fourier Transform Spectrometer (FFTS).
In an FFTS a receiver's IF   band is directly sampled
with 8/10 bit resolution using commercially available analog-to-digital
converters. Because of the high resolution, the
``spectrometer efficiency'', $\eta_{\rm S}$,
in the radiometer equation
becomes 1. This brings a factor $\approx 1.5$ improvement in
integration time compared to 2 bit/3 level autocorrelators,
for which $\eta_{\rm S}
\approx 1.23$. 
The FFT is continuously performed
with a chosen window function, which can be selected
to suppress spectral side lobes (``ringing" caused by Gibbs' phenomenon).
Thus, FFTSs have a much higher dynamic range than most other spectrometers.
In addition, FFTSs do not need total power detectors and
leveling the power levels is much simpler than in an autocorrelator. This
results in a considerable simplification of a receiver's IF section.

The FFT, as well as power spectrum averaging, is performed
in one complex logic chip, a Field-Programmable Gate Array (FPGA). Autocorrelators,
in contrast, need cascaded, custom chips, making their design and production
much more complex
and expensive.

FPGA chips have a remarkable degree of integration: \textit{Everything}, the
complete spectrometer (digital 	filters, windows, FFT, power builder, and
accumulator) is on one chip.

That the
chips are mass-produced rather than custom-designed as for
correlators, makes large bandwidth FFTSs relatively affordable. In
addition, it allows a much better reaction to the market, taking
full advantage of Moore's law.  In addition, the software for
FPGA programming can be purchased ``off the shelf''.

FFTSs afford very high channel numbers,
32768 channels per 1 GHz (in early 2006), with twice the bandwidth
and twice the number of channels for the same price (ca. 20000 kEuros)
available soon. This prize is expected to come down by a factor $>5$
for institutions who learn to program the FPGAs and design their own boards
themselves, also taking advantage of a growing demand from industry, which leads
to higher chip production rates.
The very high channel numbers, while uninteresting for extragalactic
astronomy (except for maser searches), make FFTs highly attractive
allround devices fulfilling
the needs for instantaneous coverage of wide bands. For example, they are
ideally suited for unbiased spectral line surveys of a whole IF band
(see \S \ref{linesurveys}:
At K-band (23 GHz) 1 GHz bandwidth split in 32768 channels gives
a velocity resolution of 0.40 km~s$^{-1}$, well suited for many galactic
molecular cloud sources.
\begin{figure}[!h]
\begin{center}
\includegraphics[scale=0.55]{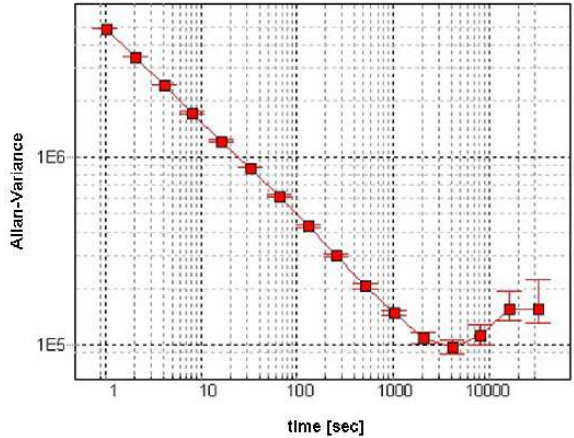}
\end{center}
\caption{Excellent stability of an MPIfR FFTS: The spectroscopic Allan time
(of ca. 4000 seconds!)
of two 1 MHz wide channels separated by 600 MHz (see Schieder et al. 1986 for a
discussion of the spectroscopic Allan variance).
The signal was provided by a temperature-stabilized noise source.}
\end{figure}
A prototypical narrow band (50 MHz), 1024 channel FFTSs, operated at the
Effelsberg 100m telescope is described by Stanko, Klein, \&\
Kerp (2005). A 500 MHz wide/16384 channel FFTS is currently in operation
at the 100m telescope.

Benz et al. (2005) describe the architecture and  building of
a 1 GHz wide/16384 channel FFTS, as well as
astronomical observations with it using the KOSMA Gornergrat telescope.
A 1 GHz bandwidth/32768 channels FFTS has been installed at the
Atacama Pathfinder Experiment (APEX) telescope
in mid-2005 and has since been performing flawlessly.
There, at 5100 m altitude,
another advantage of FFTSs, their modest power consumption, resulting
in much easier  cooling requirements, certainly adds a significant
reliability advantage over autocorrelators.

\subsection{\label{galactic}Multi-purpose usage of FFTSs}
From their concept, FFTSs lend to serilization. Once costs have come down,
making large bandwidths economically feasible, a \textit{single}
very wideband FFTS will be the spectrometer of choice for many
astronomical applications and its use will result
in \textit{huge} savings of observing time.
For an illustration let us consider two examples of interesting science
in the 3 mm band that would be enabled with an \textit{identical}
very wideband FFTS.
\subsubsection{\label{linesurveys}Unbiased spectral line surveys}
Unbiased line surveys are a powerful means to characterize the
chemistry of an astronomical source (see \S \ref{ngc253}). Starting
with the Onsala line surveys of the Orion-A star-forming region
and the star IRC+10216
(Johansson et al. 1984) the 3 mm band has been surveyed for a number of sources.
In a recent survey of the most prolific
galactic line source Sgr B2 in the 80 -- 116 GHz range with the
IRAM 30m telescope more than 4000 spectral features were found
(Belloche et al. in prep.).
Due to the use of relatively narrowband SIS receivers and
limited spectrometer capability (500 MHz wide filterbanks)
these observations required a total of 90 single spectra
of 20 minutes integration time each (allowing for some overlap).
With a wide enough FFTS and a wide bandwidth HEMT receiver the same
project would have required a single spectrum.

\subsubsection{Multi-molecule large scale mapping}
It is also worth noting that the 3 mm  range contains the ground-state
transitions of many important molecules, a.o. CO and its isotopes, HCN, CN,
HCO$^+$, C$_2$H, and the $J=2-1$ lines of CS and SiO.
Since many of these have large dipole moments and, thus, substantial
critical densities,  except for very high density/high temperature cores
near embedded protostars, only their low-$J$ lines are expected to show spatially
extended emission, Within the next few years HEMT focal plane arrays
will revolutionize single dish mm- (and cm-)wavelength astronomy.
Mapping the emission from the above
(and other) molecules, whose larger-scale distribution is poorly
known at present, will, a.o., result in the identification of star-forming regions
within molecular clouds.
Such mapping does of course not require coverage of the whole receiver
band, but only, say,  $\pm 15$ MHz around the frequency of each target
line. This means that for a 100 element array, a total usable bandwidth
of $100\times30$ MHz = 3 GHz is required. If one wanted to image 12 lines
simultaneously, one would need an FFTS with  36 GHz bandwidth, identical
to the bandwidth one would need for contiguous coverage of the 80 -- 116 GHz range
with one element as discussed above. Such mapping would yield an instantaneous
chemical fingerprint of large-scale molecular cloud regions.

\acknowledgements I am grateful to Bernd Klein for
comments on the text 
and discussions on FFTSs
(and, above all, to him and his group for building them). I would like to thank
Ian Smail for correspondence on optical identification of submm sources,
Axel Wei\ss\ and Andrew Baker for comments on the text
and
Sergio Martin for an electronic version of the NGC 253 spectrum.

\end{document}